\documentclass{PoS}
\usepackage{subfigure}
\title{Holographic light-front QCD in B meson phenomenology}

\ShortTitle{LFH QCD in B PHysics}

\author{\speaker{Mohammad Ahmady}\thanks{I would like to thank the organizers of LC2019 for making this successful event possible.}\\
        Department of Physics, Mount Allison University, Sackville, New Brunswick, Canada, E4L 1E6\\
        E-mail: \email{mahmady@mta.ca}}

\abstract{The light-front wavefunction for mesons predicted by holographic light-front QCD is used to obtain the light-cone distribution amplitudes for the light vector mesons and the transition form factors for B and $B_s$ decays to $\rho$, $K^*$ and $\phi$ mesons. Here, I present some of our predictions as compared to lattice QCD, OCD sum rules and the available experimental data.}

\FullConference{Light Cone 2019 - QCD on the light cone: from hadrons to heavy ions - LC2019\\
		16-20 September 2019\\
		Ecole Polytechnique, Palaiseau, France}

\begin{document}

\section{Introduction}
Inclusive and exclusive rare $B_{(s)}$-meson decays are important venues for testing the Standard Model (SM) and searching for signals of the new physics beyond it\cite{Ahmady:2001qh,Ahmady:2006yr}.  Dedicated experiments like LHCb at CERN and BELLE II at KEK are providing data at increasing higher precision and facilitate access to decay channels that were out of reach before.  Therefore, it is essential that the theoretical predictions for these decay modes are provided at matching accuracy with sources of uncertainties well understood.  Strong force in the nonperturbative regime is one of the main sources of theoretical uncertainty in predicting exclusive rare $B_{(s)}$-meson decays.  Observables like decay constants are directly sensitive to hadronic bound-state structure.  Moreover, the transition form factors (TFFs) which enter in the calculation of the decay rates represent the nonperturbative inputs in the theoretical predictions.  The actual computation of the form factors requires a model for strong interactions in the nonperturbative regime or through lattice QCD, which is based on QCD Lagrangian but needs significant computing power.

Traditionally, QCD Sum Rules (QCDSR) has been used to obtain the nonperturbative inputs in predicting the SM predictions on rare $B_{(s)}$ decays.  However, in recent years, an alternative approach for dealing with these nonperturbative parameters has been proposed \cite{Ahmady:2012dy,Ahmady:2013cva,Ahmady:2013cga,Ahmady:2014sva,Ahmady:2019hag}.  This new method is based on the holographic light-front QCD (HLFQCD) in which the transverse part of the valence light-front wavefunction of a meson satisfies the so-called holographic Schr\"odinger Equation (HSE) \cite{deTeramond:2008ht}
\begin{equation}
\left(-\frac{\mathrm{d}^2}{\mathrm{d}\zeta^2}-\frac{1-4L^2}{4\zeta^2} + U_{\mathrm{eff}}(\zeta) \right) \phi(\zeta)=M^2 \phi(\zeta) \;,
\label{hSE}
\end{equation}
where $M$ is the meson mass and $\mathbf{\zeta} = \sqrt{z\bar{z}} b$ ($\bar{z} = 1-z$). $b$ is the transverse separation of the quark and antiquark and $z$ is the light-front momentum fraction carried by the quark.  The confining potential is given in terms of the fundamental scale $\kappa$ and the total angular momentum $J$:
\begin{equation}
U_{eff}(\zeta)=\kappa^4 \zeta^2 + 2 \kappa^2 (J-1)\;.
\label{cp}
\end{equation}
The Regge slope for the vector mesons fixes $\kappa =0.54$ GeV\cite{Brodsky:2014yha}.  The solution of the HSE for the vector mesons (like $\rho$, $K^*$ and $\phi$), allowing for small quark masses, is written as
\begin{equation}  \Psi_{\lambda} (z,\zeta) = {\mathcal N}_{\lambda} \sqrt{z (1-z)}  \exp{ \left[ -{ \kappa^2 \zeta^2  \over 2} \right] }
\exp{ \left[ -{{(1-z)m_q^2+zm_{\bar q}^2} \over 2 \kappa^2 z(1-z) } \right]} \;.
\label{hwf}
\end{equation}
${\mathcal N}_{\lambda}$($\lambda =T,\; L$) is the polarization-dependent normalization constant and can be fixed from the requirement
 \begin{equation}
 \sum_{h,\bar{h}} \int {\mathrm d}^2 {\mathbf{r}} \, {\mathrm d} x |
 \Psi_{ \lambda} ^{h, {\bar h}}(x, r)|^{2} = 1 \,.
 \end{equation}
 The summation above is over quark and antiquark helicities after including the appropriate helicity wavefunction in (\ref{hwf})\cite{Ahmady:2016ufq}.  The quark (anti-quark) mass $m_q$ ($m_{\bar q}$) can be determined by using (\ref{hwf}) to predict the diffractive vector meson production and compare with the existing data\cite{Ahmady:2016ujw}.  In the following, I present some of the predictions for the vector meson decay constants, $B_{(s)}\to V$ ($V=\rho ,\; K^*,\; \phi$) TFFs, as well as rare B$_{(s)}$ decay rates obtained from the holographic light-front wavefunction (HLFWF).

\section{Decay constants}
Calculation of the vector meson decay constant defined as 
\begin{equation}
f_{V} P^+=\langle 0|\bar{q} (0) \gamma^{+} q(0) | V(P,L) \rangle
\label{fv-def}
\end{equation}
provides a direct test of the HLFWF \cite{Ahmady:2016ujw,Ahmady:2012dy}.  $f_V$ is directly related to the experimentally measured electronic decay width $\Gamma_{V \rightarrow e^+ e^-}$ of the vector meson:
\begin{equation}
 \Gamma_{V \rightarrow e^+ e^-}={ 4 \pi  \alpha_{em}^2  C_V^2 \over 3 M_V }f_V^2 
 \end{equation}
where $C_\phi=1/3$ for the $C_\rho=1/\sqrt{2}$.  Table \ref{tab:Decay-width} shows our predictions for the leptonic width of $\rho$ and $\phi$ for two distinct values of the quark mass versus the experimental data.  We observe that in both systems the predicted value is somewhat below the data\cite{Ahmady:2016ujw}.  One possible explanation for this discrepancy could be the lack of short-distance perturbative effects in the confining potential (\ref{cp}).
	\begin{table}
		\centering
		\begin{tabular}{|c|c|c|c|}
			\hline\hline
			Meson  &  $f_V$ [GeV] &$\Gamma_{e^+e^-}$ [KeV] & $\Gamma_{e^+e^-}$[KeV] (PDG) \\ \hline
			$\rho$ & $0.210,0.211$ & $6.355,6.383$ & $7.04 \pm 0.06$\\ \hline
			$\phi$ & $0.191,0.205$  & $0.891,1.024$  & $1.251 \pm 0.021$ \\ \hline
		\end{tabular}
		\caption{Predictions for the electronic decay widths of the $\rho$ and $\phi$ vector mesons using the HLFWF with $m_{u,d}=0.046,0.14$ GeV and $m_s=0.357,0.14$ GeV.}
		\label{tab:Decay-width}
	\end{table}
	For $K^*$, the decay constant is experimentally measured via $\Gamma(\tau^- \to K^{*-} \nu_{\tau})$ decay. In Table \ref{tab:decay}, we compare our predictions for $f_{K^*}$ and "transverse decay constant" $f_{K^*}^{\perp}$ defined as:
	\begin{equation}
	\langle 0|\bar q [\gamma^\mu,\gamma^\nu] s|K^* (P,\epsilon)\rangle =2 f_{K^*}^{\perp} (\epsilon^{\mu} P^{\nu} - \epsilon^{\nu} P^{\mu})\;,	
	\label{DA:phiperp}
	\end{equation}
	for three choices of quark mass with experimental data.
	\begin{table}[h]
		\[
		\begin{array}
		[c]{|c|c|c|c|c|c|c|}\hline
		\mbox{Approach}&\mbox{Scale}~ \mu  &m_{\bar{q}} \mbox{[MeV]} & m_s \mbox{[MeV]} &f_{K^*} \mbox{[MeV]} &f_ {K^*}^{\perp} (\mu) \mbox{[MeV]}&f_{K^*}^{\perp}/f_{K^*} (\mu)\\ \hline
		\mbox{AdS/QCD} & \sim 1~\mbox{GeV} & 140 & 280 & 200  & 118 & 0.59 \\ \hline
		\mbox{AdS/QCD} & \sim 1~\mbox{GeV} & 195 & 300 & 200  & 132 & 0.66 \\ \hline
		\mbox{AdS/QCD} & \sim 1~\mbox{GeV} & 250 & 320 & 200  & 142 & 0.71 \\ \hline \mbox{Experiment}  & &  & &205\pm 6  & & \\ \hline
		\mbox{Lattice}  & 2 ~\mbox{GeV}& & & & &0.780 \pm 0.008 \\ \hline
		\mbox{Lattice}  & 2 ~\mbox{GeV}& &  & &  & 0.74 \pm 0.02\\ \hline
		\end{array}
		\]
		\caption{Comparison between AdS/QCD predictions for the decay constant of the $K^*$ meson with experiment (obtained from $\Gamma(\tau^- \to K^{*-} \nu_{\tau})$), and the ratio of couplings with lattice data.}
		\label{tab:decay}
	\end{table}
	Our results indicate that, unlike $f_{K^*}$, $f_{K^*}^{\perp}$ is sensitive to the choice of $m_{\bar q}$ and $m_s$.  However, it should be pointed out that better agreement with lattice data on the ratio $f_{K^*}^{\perp}/f_{K^*}$ is achieved when constituent quark masses are used\cite{Ahmady:2012dy}.
	
	\section{Light-cone distribution amplitudes}
\begin{figure}
			\centering
			\subfigure{\includegraphics[width=.40\textwidth]{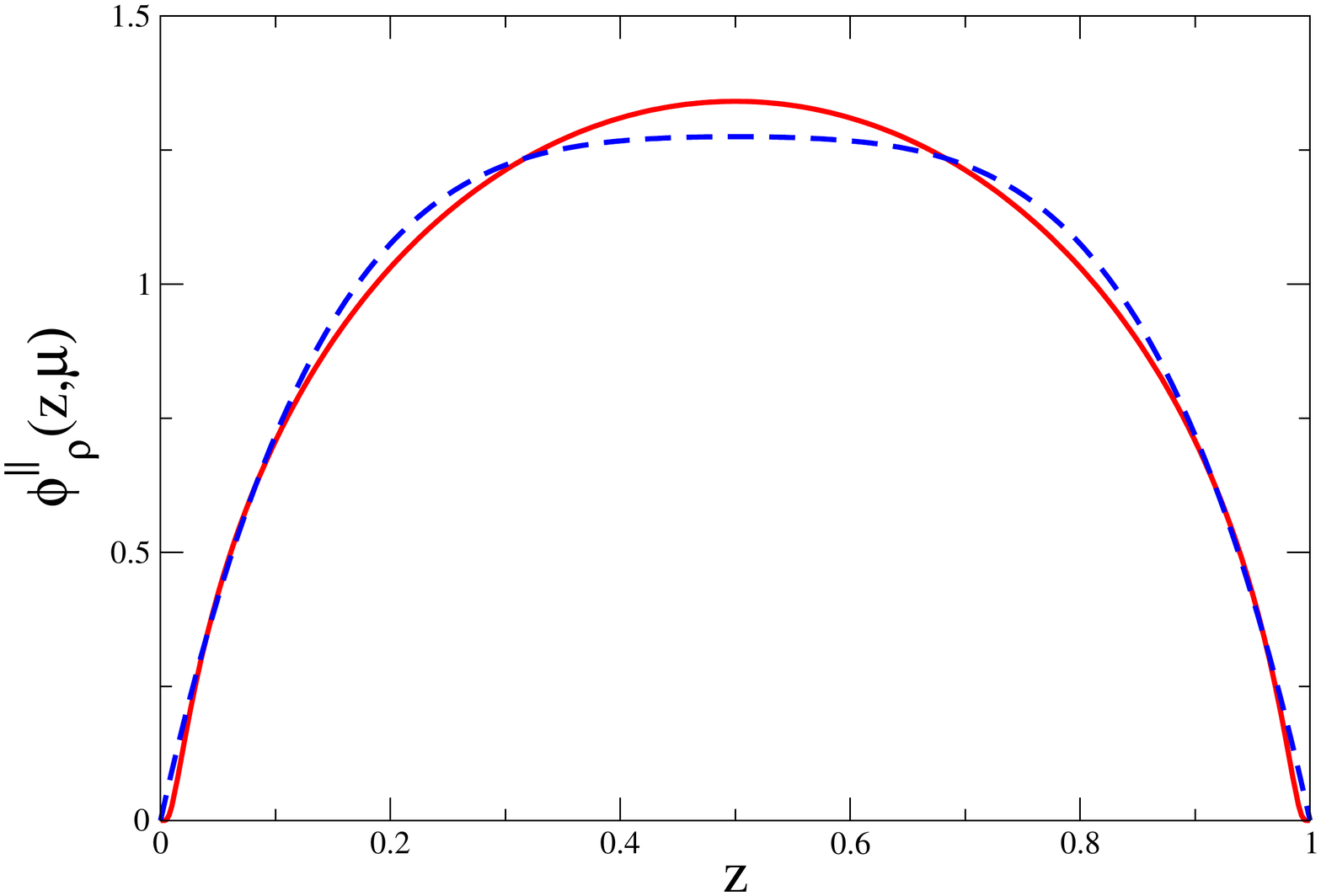} }
			\subfigure{\includegraphics[width=.40\textwidth]{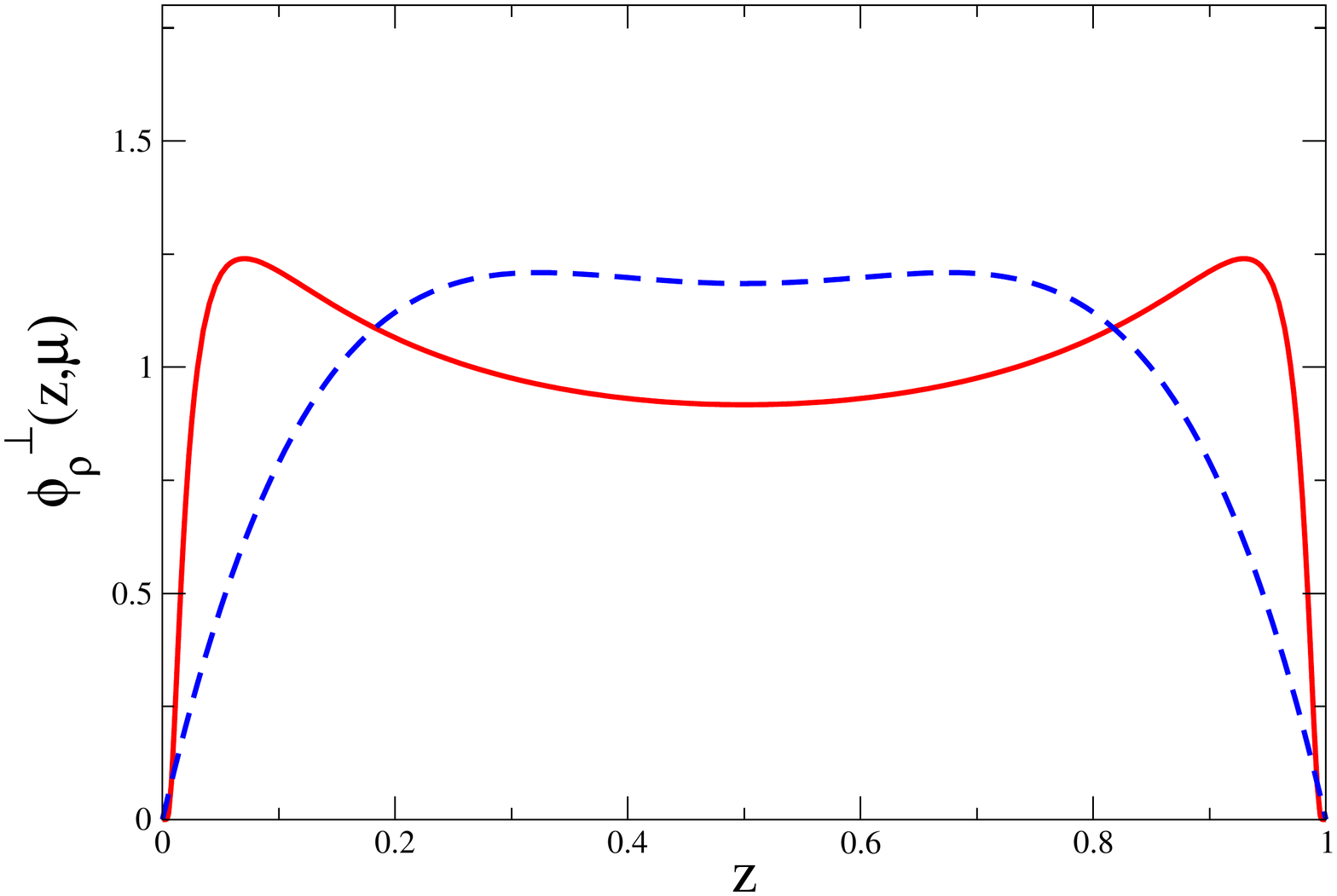} }
			\caption{Twist-$2$ LCDAs for the $\rho$ meson. Solid Red: HLFQCD DAs; Dashed Blue: Sum Rules DAs.} \label{fig:tw2DAs}
		\end{figure}

\begin{figure}[htbp]
				\begin{subfigure}{}
					\centering
					\includegraphics[width=0.4\textwidth]{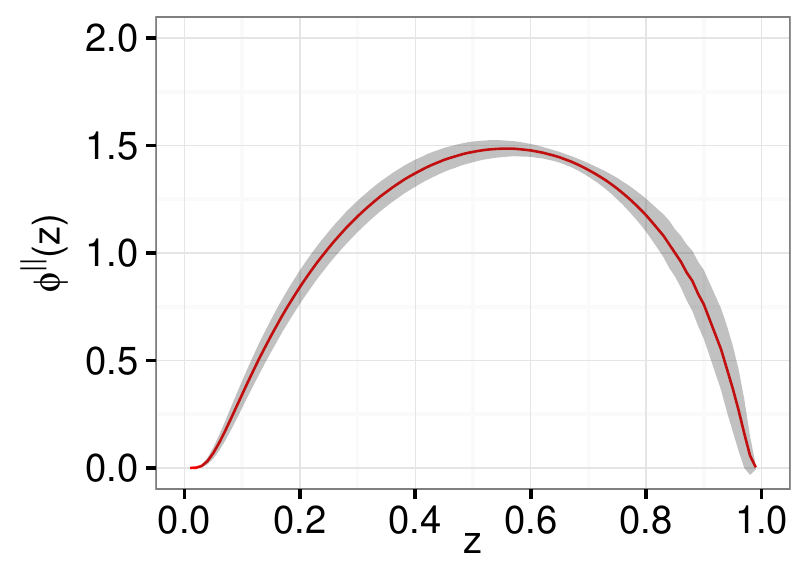}
					\label{adsdapara}
				\end{subfigure}
				\hspace{.1cm}
				\begin{subfigure}{}
					\centering
					\includegraphics[width=0.4\textwidth]{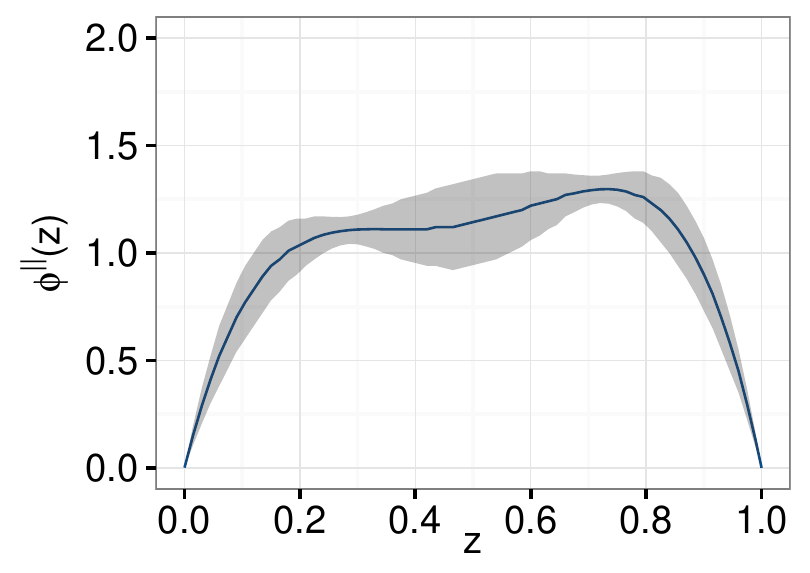}
					\label{srdapara}
				\end{subfigure}\\
				\begin{subfigure}{}
					\centering
					\includegraphics[width=0.4\textwidth]{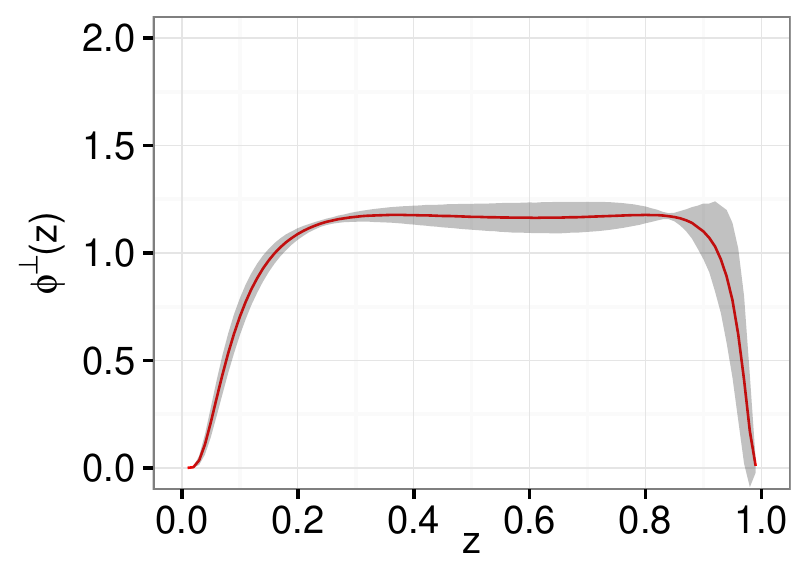}
					\label{adsdaperp}
				\end{subfigure}
				\begin{subfigure}{}
					\centering
					\includegraphics[width=0.4\textwidth]{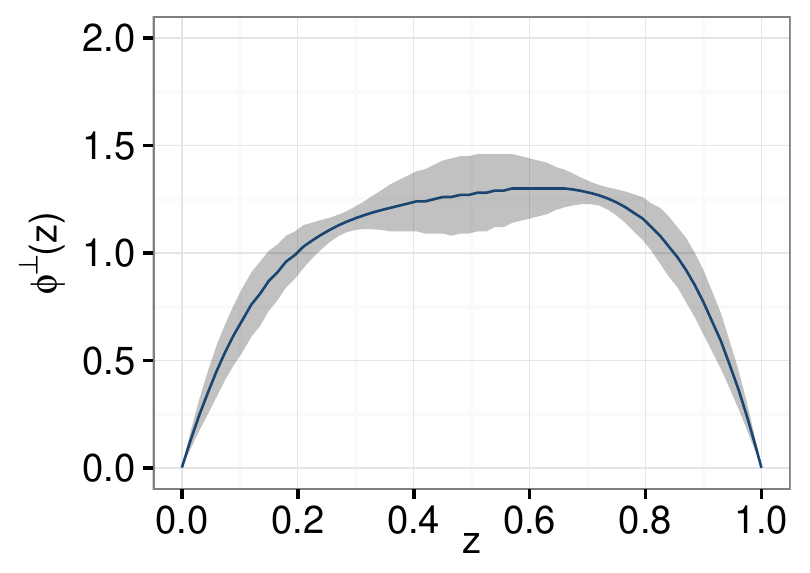}
					\label{srperp}
				\end{subfigure}
				\caption{Twist-2 LCDAs for $K^*$ predicted by HLFQCD (graphs on the left) and SRQCD (graphs on the right).  The uncertainty bands are due to the variation of the quark masses for HLFQCD, and the error bar on Gegenbauer coefficients for SRQCD predictions.}
				\label{das}
			\end{figure}

	The light-cone distribution amplitudes (LCDAs) are nonperturbative universal functions which are obtained from light-front wavefunction \cite{Ahmady:2013cva}.  In figure \ref{fig:tw2DAs}, we compare twist-2 LCDAs for the $\rho$ meson with those obtained from QCDSR.
		For the $K^*$ meson, the same comparison is made in figure \ref{das} including the respective uncertainty bands due to the variation in the quark mass in HLFQCD as opposed to error bar on Gegenbauer coefficients in SRQCD\cite{Ahmady:2018fvo}.  LCDAs directly enter the calculation of the observables which involve the spectator quark in a meson, like isospin asymmetry in $B\to K^*\gamma$ decay\cite{Ahmady:2013cva}. Indeed, one can show that the isospin asymmetry in this particular rare decay can be expressed in terms of the integrals of LCDAs.  It turns out that the QCDSR prediction for one of these integrals diverges due to the end point behavior of $\phi^\perp_{K^*}$.  HLFQCD DAs, on the other hand, have no such divergence ambiguity and in fact the predicted value for the asymmetry, i.e. 3.3\%, is in agreement with the data within the error bars\cite{Ahmady:2013cva}.  
			
			\section{Prediction for the differential decay rates}
\begin{figure}
				\subfigure{\includegraphics[width=.40\textwidth]{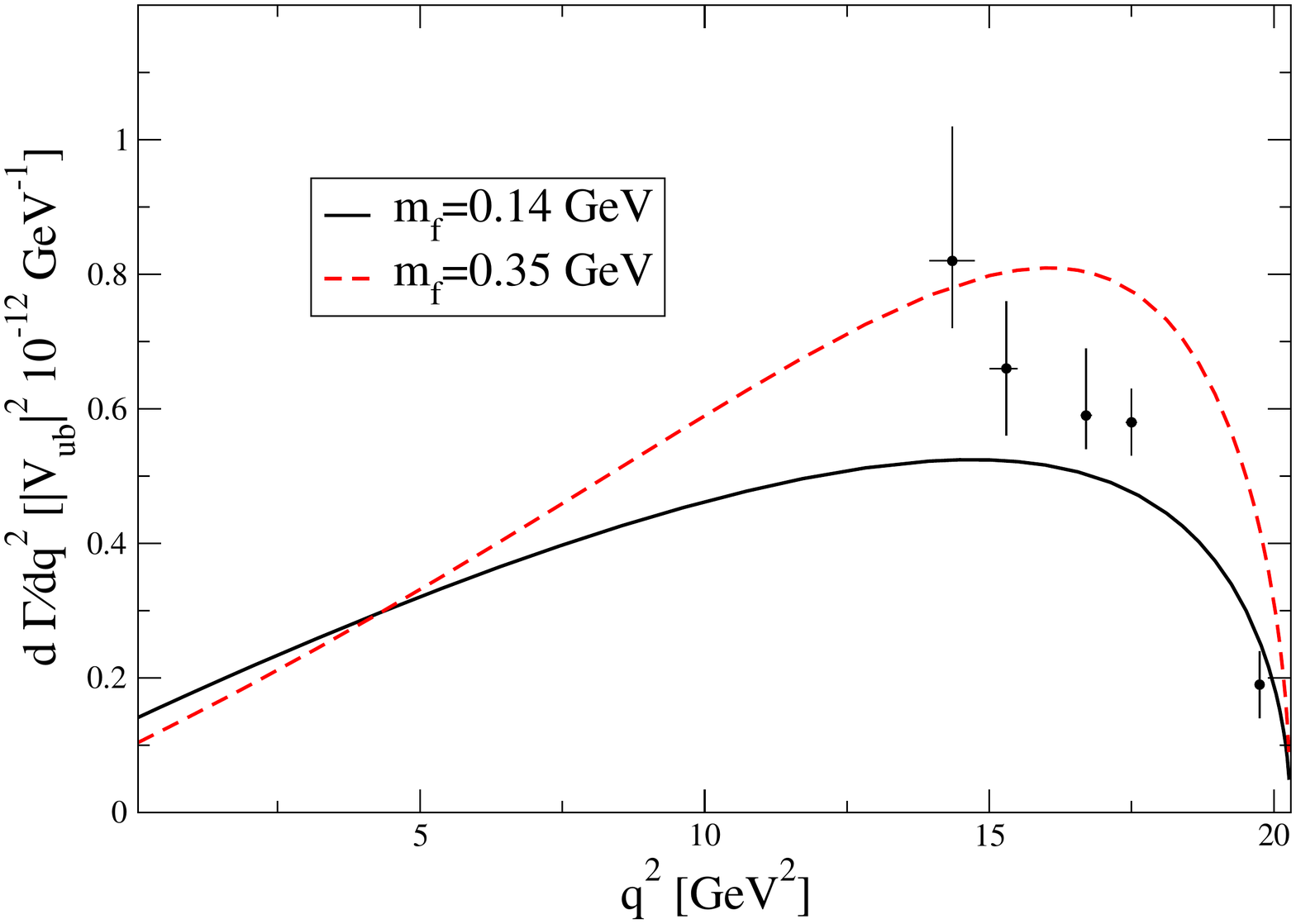} }
				\hspace{1cm}
				\subfigure{\includegraphics[width=.40\textwidth]{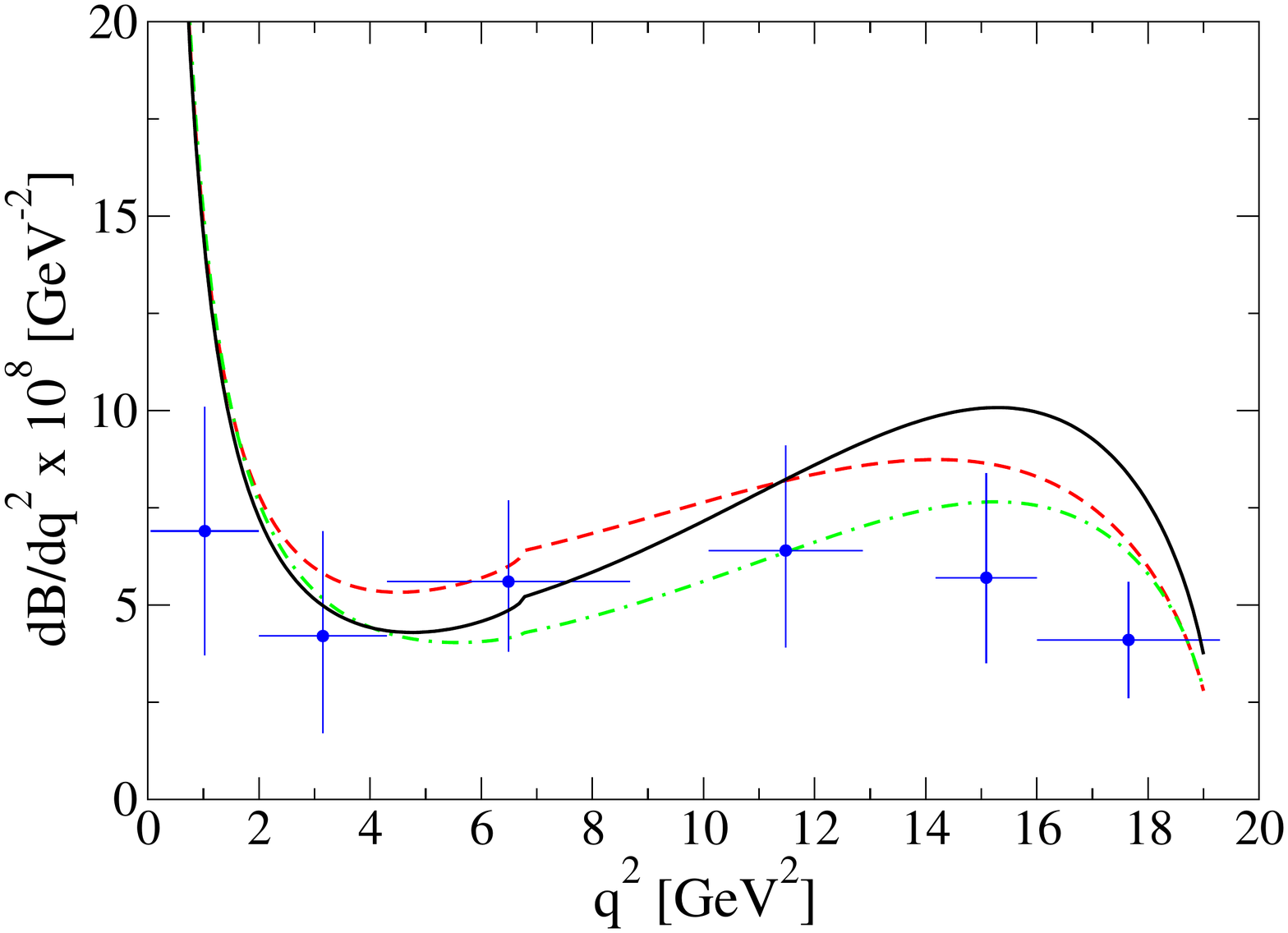} }
				\caption{Left: Differential decay rate for the semileptonic $B\to\rho\ell\bar\nu$ decay. The lattice data points are from UKQCD Collaboration. Right: Differential branching ratio for $B\to K^*\mu^+\mu^-$ decay with (solid black), without (dashed red) fitting to lattice points available at hight $q^2$.  Dashed green prediction includes new physics contributions to the Wilson Coefficient $C_9$.  The data points are from LHCb. } \label{fig:BR1}
			\end{figure}

			Transition form factors (TFFs), which enter the calculation of the differential decay rates, parametrize the hadronic matrix elements of the underlying quark decays.  These nonperturbative functions are ideally computed directly from the light-front wavefunctions of the mesons.  We are currently working on developing this method.  The alternative existing method is to use the light-cone sum rules (LCSR) which leads to expressions of the TFFs in terms of the LCDAs and decay constants\cite{Ahmady:2013cva,Ahmady:2014sva}.  Figure \ref{fig:BR1} shows the HLFQCD predictions for $B\to\rho\ell\bar{\nu}$ and $B\to K^*\mu^+\mu^-$ differential branching ratios\cite{Ahmady:2013cva}.  The experimental data points for the latter are obtained from averaging the LHCb measurements of the $B^+\to K^{*+}\mu^+\mu^-$ and $B^\circ \to K^{*\circ}\mu^+\mu^-$ decays\cite{Ahmady:2014sva}.
			
\begin{figure}
				\subfigure{\includegraphics[width=.40\textwidth]{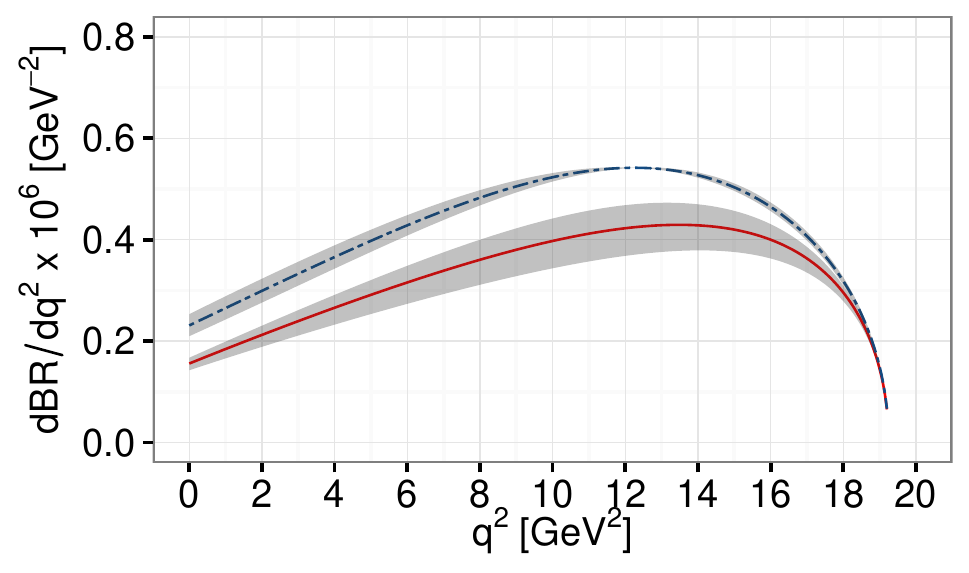} }
				\subfigure{\includegraphics[trim=1cm 15cm 3.5cm 2cm, clip, width=.40\textwidth]{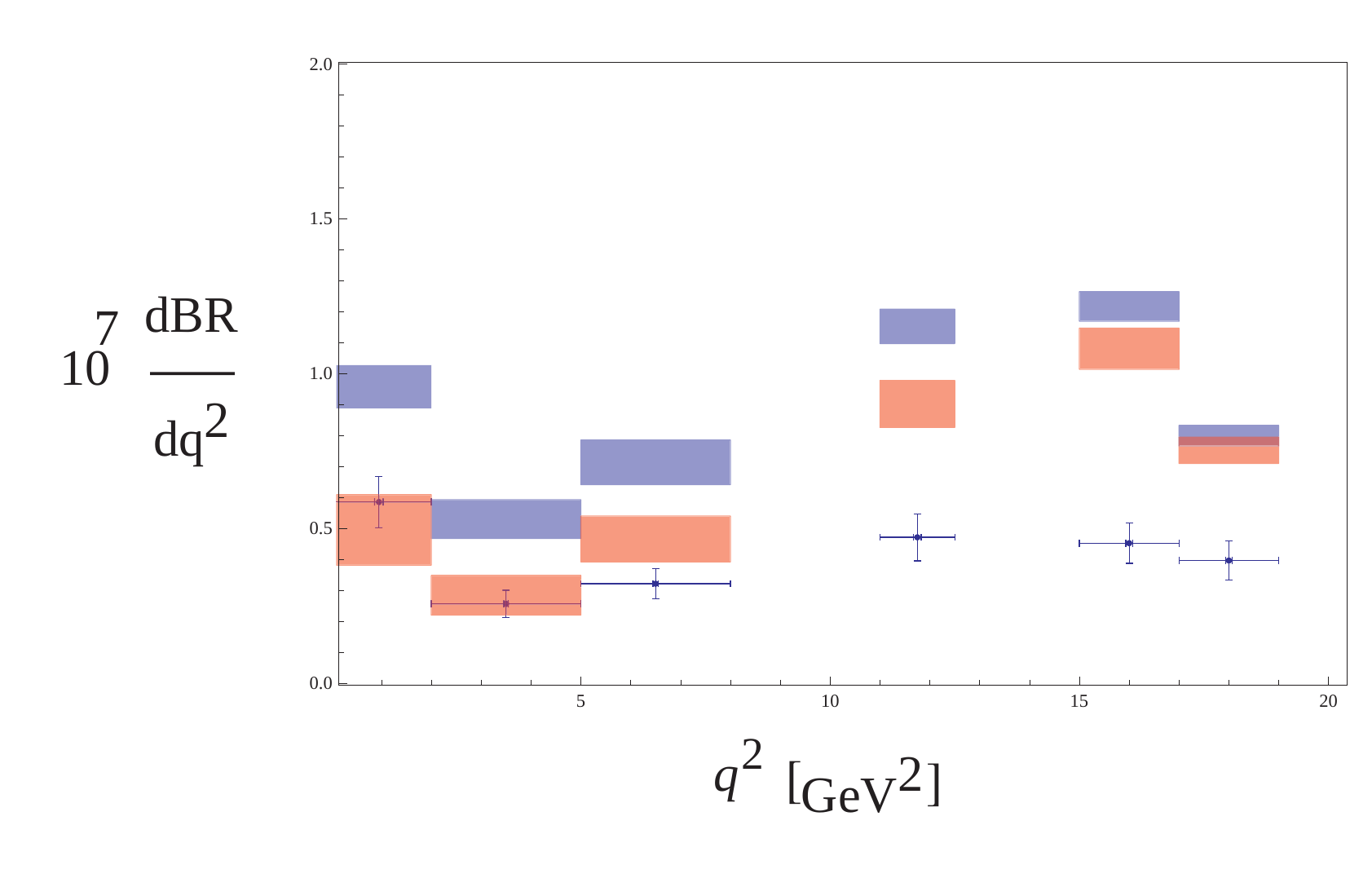} }
				\caption{Left: The HLFQCD (Solid line) and QCDSR (Dashed line) predictions for the differential Branching Ratio for $B\to K^*\nu\bar\nu$.  The shaded band represents the uncertainty coming from the form factors. Right: The differential branching ratio for $B_s\to\phi\mu^+\mu^-$ predicted by HLFQCD (red rectangles) and QCDSR (blue rectangles).  The uncertainty widths are due to the form factors. The experimental data points are measured by LHCb .} \label{fig:BR2}
			\end{figure}

			A comparison between HLFQCD and QCDSR predictions for $B\to K^*\nu\bar \nu$ and $B_s\to\phi\mu^+\mu^-$ rare decays are depicted in figure \ref{fig:BR2}.  The future measurement of the former decay at low to intermediate $q^2$ can discriminate between the two predictions\cite{Ahmady:2018fvo}.  The latter decay, on the other hand, has been measured by LHCb and we note that the HLFQCD prediction for the differential decay rate is in better agreement with the data especially at low $q^2$ range\cite{Ahmady:2019hag}. 
			
\section{conclusion}
HLFQCD provides an alternative method in calculating the nonperturbative inputs like DAs and TFFs in the theoretical computation of rare B decays.  In general, HLFQCD predictions for the differential decays rates are lower than QCDSR results and in better agremment with the available data for both $B\to K^*\mu^+\mu^-$ and $B_s\to\phi\mu^+\mu^-$ decays specifically at small $q^2$ range.

\end{document}